\documentclass[sigconf]{acmart}
\usepackage{booktabs} 
\usepackage{amsmath}
\usepackage{amsfonts}
\usepackage{comment}
\usepackage{footmisc}
\usepackage{mathrsfs}
\usepackage{amssymb}
\usepackage{graphicx}
\usepackage{subfigure}
\usepackage{multirow}
\usepackage{algorithm}
\usepackage{algorithmic}
\usepackage{multicol}  
\usepackage{enumitem}
\usepackage{array}
\usepackage{float}
\usepackage{hyperref}
\usepackage{units}
\AtBeginDocument{%
	\providecommand\BibTeX{{%
			\normalfont B\kern-0.5em{\scshape i\kern-0.25em b}\kern-0.8em\TeX}}}

\begin{document}
	\title{AutoEmb: Automated Embedding Dimensionality Search\\in Streaming Recommendations}
	\author{Xiangyu Zhao}
	\affiliation{
		\institution{Michigan State University}
	}
	\email{zhaoxi35@msu.edu}
	
	\author{Chong Wang}
	\affiliation{
		\institution{Bytedance}
	}
	\email{chong.wang@bytedance.com}

	\author{Ming Chen}
	\affiliation{
		\institution{Bytedance}
	}
	\email{ming.chen@bytedance.com}

	\author{Xudong Zheng}
	\affiliation{
		\institution{Bytedance}
	}
	\email{zhengxudong.alpha@bytedance.com}
	
	\author{Xiaobing Liu}
	\affiliation{
		\institution{Bytedance}
	}
	\email{will.liu@bytedance.com}
	
	\author{Jiliang Tang}
	\affiliation{
		\institution{Michigan State University}
	}
\email{tangjili@msu.edu}
\renewcommand{\shortauthors}{Xiangyu Zhao et al.}
	
	\begin{abstract}
		Deep learning based recommender systems (DLRSs) often have embedding layers, which are utilized to lessen the dimensionality of categorical variables (e.g. user/item identifiers) and meaningfully transform them in the low-dimensional space. The majority of existing DLRSs empirically pre-define a fixed and unified dimension for all user/item embeddings. It is evident from recent researches that different embedding sizes are highly desired for different users/items according to their popularity. However, manually selecting embedding sizes in recommender systems can be very challenging due to the large number of users/items and the dynamic nature of their popularity. Thus, in this paper, we propose an AutoML based end-to-end framework (AutoEmb), which can enable various embedding dimensions according to the popularity in an automated and dynamic manner. To be specific, we first enhance a typical DLRS to allow various embedding dimensions; then we propose an end-to-end differentiable framework that can automatically select different embedding dimensions according to user/item popularity; finally we propose an AutoML based optimization algorithm in a streaming recommendation setting. The experimental results based on widely used benchmark datasets demonstrate the effectiveness of the AutoEmb framework.
	\end{abstract}
	
	\maketitle
	\begin{figure}
	\centering
	\includegraphics[width=81mm]{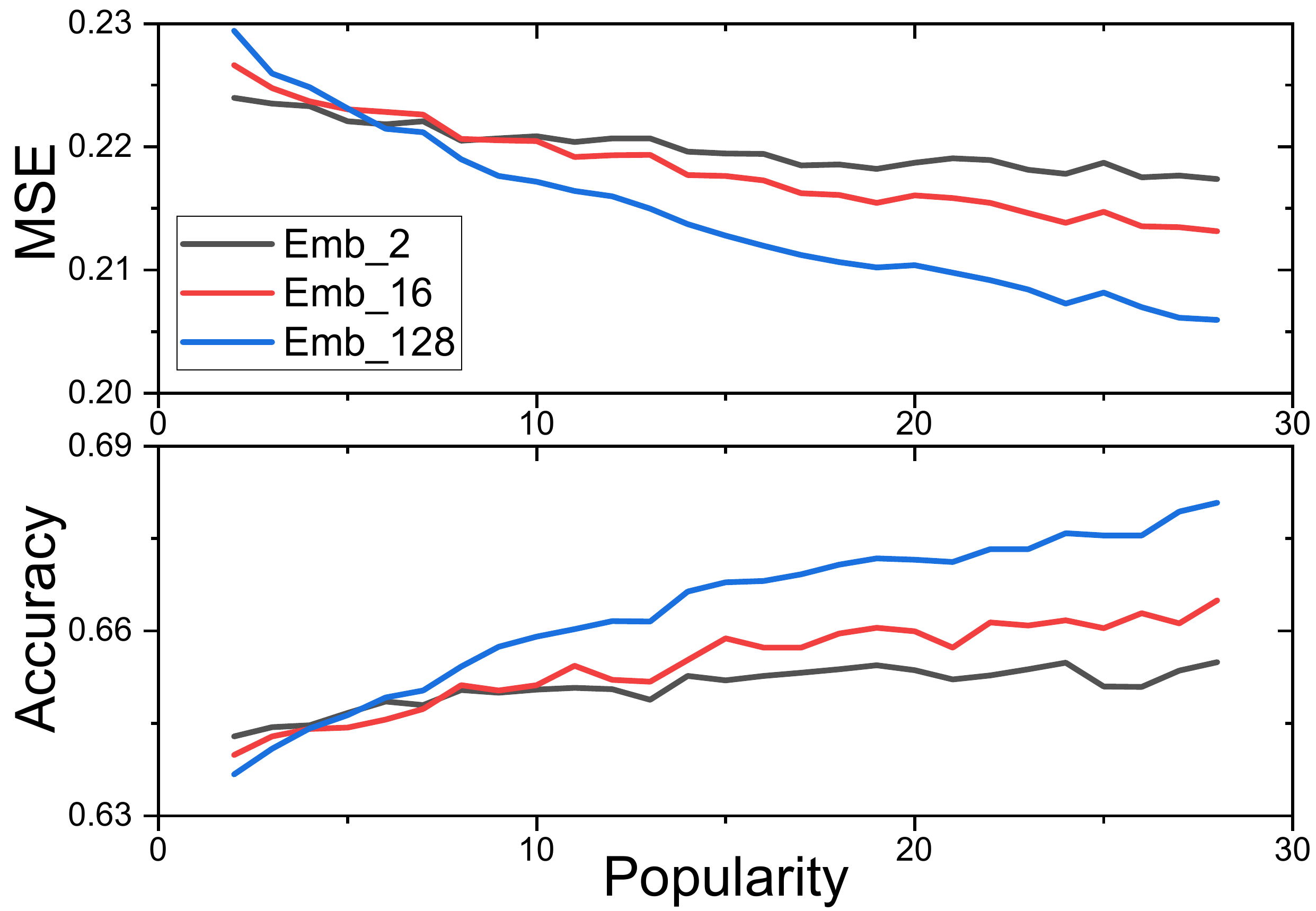}
	\caption{Impact of embedding dimensionality on recommendations.}
	\label{fig:Fig1_example}
\end{figure}
\section{Introduction}
\label{sec:introduction}
Driven by the recent advances in deep learning, there have been increasing interests in developing deep learning based recommender systems (DLRSs)~\cite{zhang2017deep,nguyen2017personalized,wu2016personal}. DLRSs have boosted the recommendation performance because of their capacity of effectively catching the non-linear user-item relationships, and learning the complex abstractions as data representations~\cite{zhang2019deep}. Architectures of DLRS often mainly consist of three key components: (i) \textit{embedding layers} that map raw user/items features in a high dimensional space to dense vectors in a low dimensional \textit{embedding space}, (ii) \textit{hidden layers} that perform nonlinear transformations to transform the input features, and (iii) \textit{output layers} that make predictions for specific recommendation tasks (e.g. regression and classification) based on the representations from hidden layers. The majority of existing researches have focused on designing sophisticated neural network architectures for the hidden layers and output layers, while the embedding layers have not gained much attention. However, in the large-scale real-world recommender systems with numerous users and items, embedding layers play a tremendously crucial role in accurate recommendations. The most typical use of embedding is to transform an identifier, i.e., user-id or item-id, into a real-valued vector. Each embedding can be considered as the latent representation of a specific user or item. Compared to hand-crafted features, well-learned embeddings have been demonstrated to significantly enhance the recommendation performance~\cite{cheng2016wide,guo2017deepfm,pan2018field,qu2016product,zhou2018deep}. This is because embeddings can reduce the dimensionality of categorical variables (e.g. one-hot identifiers) and meaningfully represent users/items in the latent space. Furthermore, nearest neighbors in the embedding space can be viewed as similar users/items; while the mapping of one-hot space is completely uninformed where similar users/items are not projected closer to each other.

The majority of existing DLRSs often adopt a unified and fixed dimensionality in their embedding layers. In other words, all users (or items) share the same and fixed embedding size. It naturally raises a question -- do we need different embedding dimensions for different users/items?  To investigate this question, we conduct a preliminary study on the movielens-20m dataset~\footnote{https://grouplens.org/datasets/movielens/20m/}. For each user, we first select a fixed part of his/her \textit{ratings} (labeled interactions with items) as test and then we choose $x$ ratings as training.  Figure~\ref{fig:Fig1_example} illustrates how the recommendation performance of a typical DLRS~\cite{cheng2016wide} with embedding dimensions $2$, $16$ and $128$ in terms of the mean-squared-error (MSE) and accuracy changes when we vary $x$.  Lower MSE (or higher accuracy) means better performance. Note that we refer to the number of interactions users/items have as popularity in this work. From the figure, with the increasing of the popularity $x$, (i) the performance of models with different embedding dimensions increases but larger embedding dimensions gain more; and (ii) smaller embedding dimensions first work better and then are outperformed by larger embedding dimensions. These observations are quite expected since the embedding size often determines the number of model parameters to learn and the capacity to encode information by the embedding. On the one hand, smaller embedding dimensions often mean fewer model parameters and lower capacity. Thus, they can work well when the popularity is small. However, the capacity limits the performance with the increasing popularity when the embedding needs to encode more information. On the other hand, larger embedding dimensions usually indicate more model parameters and higher capacity. They typically need sufficient data to be well trained. Therefore they cannot work well when the popularity is small but they have the potential to capture more information as the popularity increases.  Given that users/items have very different popularity in a recommender system, different embedding dimensions should be allowed by DLRSs. This property is highly desired in practice since real-world recommender systems are streaming where the popularity is highly dynamic. For example, new interactions are rapidly occurred and new users/items are continuously added. 


In this paper, we aim to enable different embedding dimensions for different users/items at the embedding layer under the streaming setting. We face tremendous challenges. First, the number of users/items in real-world recommender systems is very large and the popularity is highly dynamics, it is hard, if possible, to manually select different dimensions for different users/items. Second, the input dimension of the first hidden layer in existing DLRSs is often unified and fixed, it is difficult for them to accept different dimensions from the embedding layers. Our attempt to solve these challenges leads to an end-to-end differentiable AutoML based framework (AutoEmb), which can make use of various embedding dimensions in an automated and dynamic manner. Our experiments in real-world e-commerce data demonstrate the effectiveness of the proposed framework. 

	\begin{figure}[t]
	\centering
	\includegraphics[width=76mm]{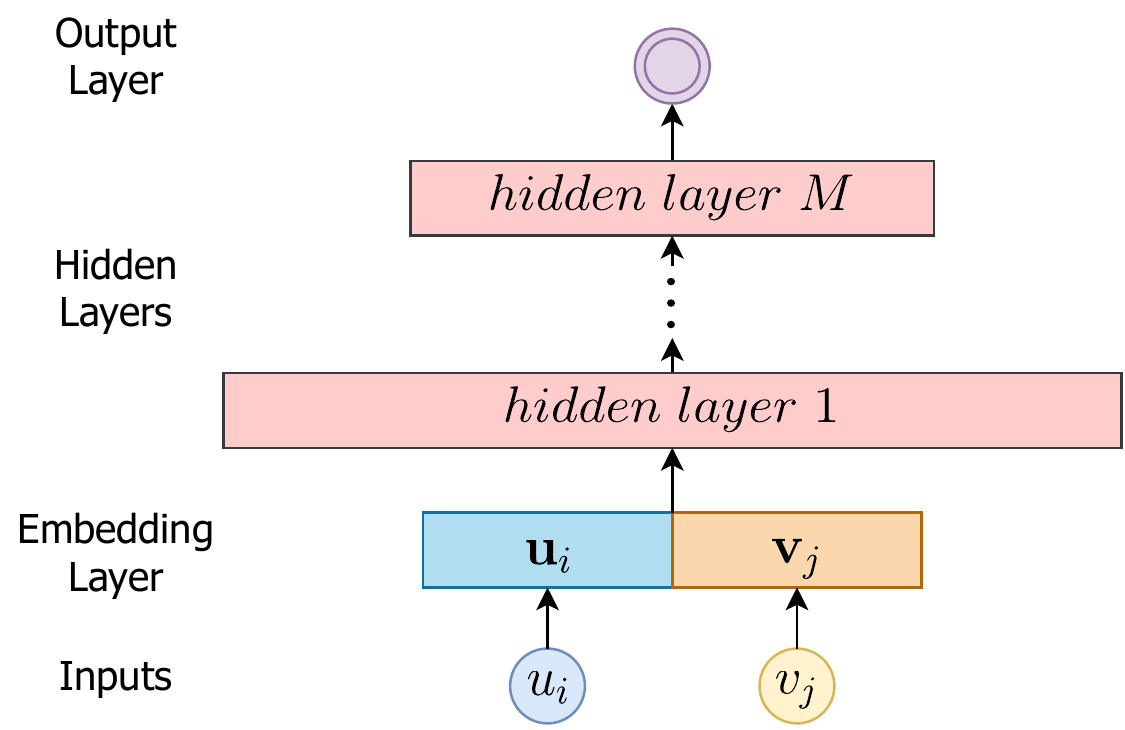}
	\caption{The basic DLRS architecture.}
	\label{fig:Fig2_DLRS}
\end{figure}
\section{Framework}
\label{sec:framework}
As discussed in Section~\ref{sec:introduction}, assigning different embedding sizes to different users/items in the streaming recommendation setting faces two challenges. To address these challenges, we propose an AutoML based end-to-end framework, which can automatically and dynamically leverage embedding with various dimensions. In the following, we will first illustrate a widely used DLRS as our basic architecture, then we will enhance it to enable various embedding dimensions, next we will discuss how to automatically and dynamically select the various dimensions, and finally, an AutoML based optimization algorithm will be provided for the streaming recommendation.

\subsection{A Basic DLRS Architecture}
\label{sec:DLRS}
We illustrate a basic DLRS architecture in Figure~\ref{fig:Fig2_DLRS}, which contains three components: (i) the embedding layers that map user/item IDs ($u_i,v_j$) into dense and continuous valued embedding vectors ($\mathbf{u}_i,\mathbf{v}_j$), (ii) the hidden layers which are fully connected layers that non-linearly transform the embedding vectors ($\mathbf{e}_i,\mathbf{e}_j$) into hierarchical feature representations, and (iii) the output layer that generates the prediction for recommendation. Given a user-item interaction, the DLRS first performs embedding-lookup processes according to the user-id and item-id, and concatenates the two embeddings; then the DLRS feeds the concatenated embedding and makes predictions via the hidden and output layers. This typical DLRS architecture is widely used in recent recommendations~\cite{cheng2016wide}. However, it has fixed neural network architectures, which cannot handle different embedding dimensions.  Next, we will enhance this basic DLRS architecture to enable various embedding dimensions.

\subsection{The Enhanced DLRS Model}
\label{sec:AugmentedDLRS}
As discussed in Section~\ref{sec:introduction}, shorter embeddings with fewer model parameters can generate better recommendations when the popularity is small; while with the increase of popularity, longer embeddings with more model parameters and higher capacity achieve better recommendation performance. Motivated by this observation, assigning different embedding dimensions for users/items with different popularity is highly desired. However, the basic DLRS architecture in Section~\ref{sec:DLRS} is not able to handle various embedding dimensions because of its fixed neural network architecture. 

\begin{figure}[t]
	\centering
	\includegraphics[width=85mm]{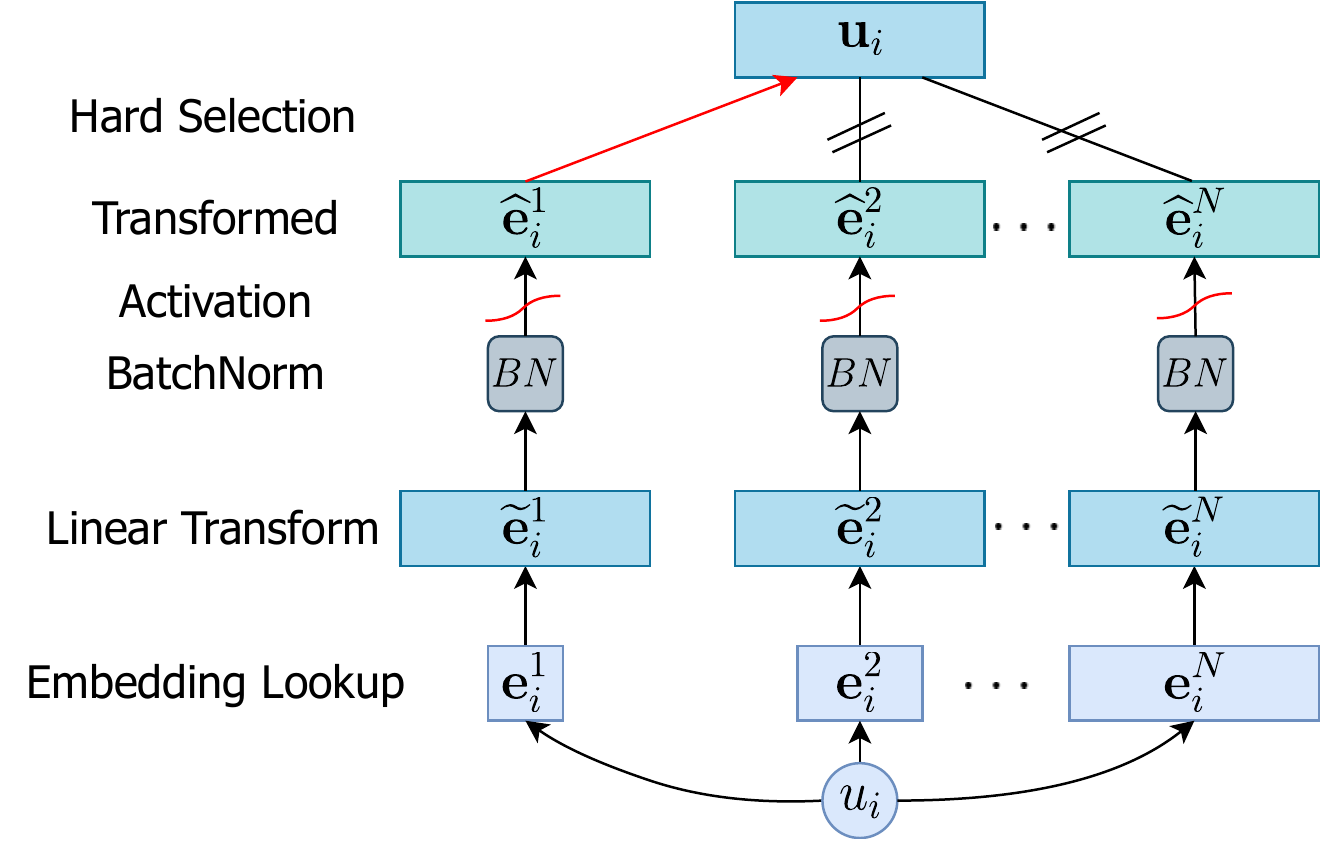}
	\caption{The embedding transformation and selection.}
	\label{fig:Fig3_Hard}
\end{figure}

The basic idea to address this challenge is to transform various embedding dimensions into the same dimension, so that the DLRS can select one of the transformed embeddings according to current user/item popularity. Figure~\ref{fig:Fig3_Hard} illustrates the embedding transformation and selection process. Suppose we have $N$ embedding spaces $\{\mathbf{E}^1,\cdots,\mathbf{E}^N\}$, and the dimension of an embedding in each space is $d_1,\cdots,d_N$, where  $d_1\textless\cdots \textless d_N$. We define $\{\mathbf{e}_i^1,\cdots,\mathbf{e}_i^N\}$ is the set of embeddings for a given user $u_i$ from all embedding spaces. To unify the embeddings vectors $\{\mathbf{e}_i^1,\cdots,\mathbf{e}_i^N\}$, we introduce a component with $N$ fully-connected layers, which transform $\{\mathbf{e}_i^1,\cdots,\mathbf{e}_i^N\}$ into same dimension $d_N$: 
	\begin{equation}
	\label{equ:Linear}
	\begin{array}{l}
	\,\,{\mathbf{\widetilde{e}}_i^1\leftarrow\mathbf{W}_{1}^{\top} \mathbf{e}_i^1+\mathbf{b}_{1}} \\\\
	\,\,{\mathbf{\widetilde{e}}_i^2\leftarrow\mathbf{W}_{2}^{\top} \mathbf{e}_i^2+\mathbf{b}_{2}} \\
	{\,\,\,\,\,\,\,\,\,\,\,\vdots} \\
	{\mathbf{\widetilde{e}}_i^N\leftarrow\mathbf{W}_{N}^{\top} \mathbf{e}_i^N+\mathbf{b}_{N}} \\
	\end{array}
	\end{equation}
\noindent where $\mathbf{W}_{1}\in \mathbb{R}^{d_1 \times d_N},\mathbf{W}_{2}\in \mathbb{R}^{d_2 \times d_N},\cdots,\mathbf{W}_{N}\in \mathbb{R}^{d_N \times d_N}$ are weight matrices and $\{\mathbf{b}_{1},\cdots,\mathbf{b}_{N}\}\in \mathbb{R}^{d_N}$ are bias vectors. After the linear transformations, we have mapped the original embedding vectors $\{\mathbf{e}_i^1,\cdots,\mathbf{e}_i^N\}$ into the same dimension $\{\mathbf{\widetilde{e}}_i^1,\cdots,\mathbf{\widetilde{e}}_i^N\}\in \mathbb{R}^{d_N}$. In practice, we can observe that the magnitude of the transformed embeddings $\{\mathbf{\widetilde{e}}_i^1,\cdots,\mathbf{\widetilde{e}}_i^N\}$ varies significantly, which makes them become incomparable. To tackle this challenge, we conduct BatchNorm~\cite{ioffe2015batch} with the Tanh activation~\cite{karlik2011performance} on the transformed embeddings $\{\mathbf{\widetilde{e}}_i^1,\cdots,\mathbf{\widetilde{e}}_i^N\}$ as follows:
	\begin{equation}
	\label{equ:BatchNorm}
	\arraycolsep=1.2pt\def\arraystretch{1.6}
	\begin{array}{l}
	\,\,\mathbf{\widehat{e}}_i^1 \leftarrow tanh(\frac{\mathbf{\widetilde{e}}_i^1-\mu_{\mathcal{B}}^1}{\sqrt{(\sigma_{\mathcal{B}}^{1})^2+\epsilon}}) \\
	\,\,\mathbf{\widehat{e}}_i^2 \leftarrow tanh(\frac{\mathbf{\widetilde{e}}_i^2-\mu_{\mathcal{B}}^2}{\sqrt{(\sigma_{\mathcal{B}}^{2})^2+\epsilon}}) \\
	{\,\,\,\,\,\,\,\,\,\,\,\vdots} \\
	\mathbf{\widehat{e}}_i^N \leftarrow tanh(\frac{\mathbf{\widetilde{e}}_i^2-\mu_{\mathcal{B}}^N}{\sqrt{(\sigma_{\mathcal{B}}^{N})^2+\epsilon}})
	\end{array}
	\end{equation}
\noindent where $\mu_{\mathcal{B}}^n$ is the mini-batch mean and $(\sigma_{\mathcal{B}}^{n})^2$ is the mini-batch variance for $\forall n \in [1,N]$. $\epsilon$ is a constant
added to the mini-batch variance for numerical stability. the Tanh function activates the normalized embedding into $(0,1)$. After BatchNorm and activation, the linearly transformed embeddings $\{\mathbf{\widetilde{e}}_i^1,\cdots,\mathbf{\widetilde{e}}_i^N\}$ become to magnitude-comparable embedding vectors $\{\mathbf{\widehat{e}}_i^1,\cdots,\mathbf{\widehat{e}}_i^N\}$ with the same dimension. Given an item $v_j$, we conduct the same transformations as these in Equations~(\ref{equ:Linear}) and~(\ref{equ:BatchNorm}) on its embeddings $\{\mathbf{f}_j^1,\cdots,\mathbf{f}_j^N\}$, and obtain magnitude-comparable ones $\{\mathbf{\widehat{f}}_j^1,\cdots,\mathbf{\widehat{f}}_j^N\}$ that share the same dimension. 

\begin{figure}[t]
	\centering
	\includegraphics[width=81mm]{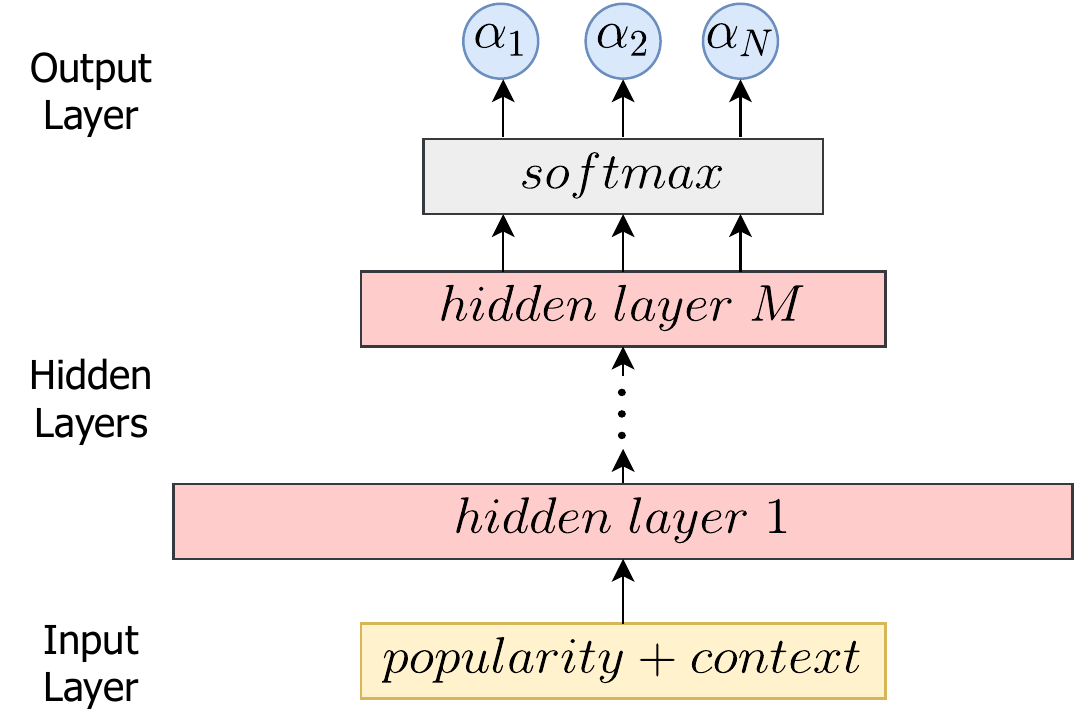}
	\caption{The controller architecture.}
	\label{fig:Fig5_controller}
\end{figure}

According to the popularity, the DLRS will select a pair of transformed embeddings $(\,\mathbf{\widehat{e}}_i^{\,*}; \mathbf{\widehat{f}}_j^{\,*} \,)$ as the representations of the user $u_i$ and item $v_j$:
	\begin{equation}
	\label{equ:Hardselection}
	\arraycolsep=1.2pt\def\arraystretch{1.6}
	\begin{array}{l}
	\mathbf{u}_i = \mathbf{\widehat{e}}_i^{\,*},\;\;\; where \;\;\mathbf{\widehat{e}}_i^{\,*}\in \{\mathbf{\widehat{e}}_i^1,\cdots,\mathbf{\widehat{e}}_i^N\}\\
	\mathbf{v}_j = \mathbf{\widehat{f}}_j^{\,*},\;\;\; where \;\; \mathbf{\widehat{f}}_j^{\,*} \in \{\mathbf{\widehat{f}}_j^1,\cdots,\mathbf{\widehat{f}}_j^N\}
	\end{array}
	\end{equation}
The embedding size is selected by a controller that will be detailed in the next subsection. Then, we concatenate user's and item's representations $\mathbf{h}_{0}=\left[\mathbf{u}_i;\mathbf{v}_j\right]$ and feed $\mathbf{h}_{0}$ as the input into $M$ fully-connected hidden layers:
	\begin{equation}
	\begin{array}{l}
	{\,\,\mathbf{h}_{1}=tanh\left(\mathbf{W}_{1}^{\top} \mathbf{h}_{0}+\mathbf{b}_{1}\right)} \\
	{\,\,\,\,\,\,\,\,\,\,\,\vdots} \\
	{\mathbf{h}_{M}=tanh\left(\mathbf{W}_{M}^{\top} \mathbf{h}_{M-1}+\mathbf{b}_{M}\right)} 
	\end{array}
	\end{equation}
\noindent where $\mathbf{W}_{m}$ is the weight matrix and $\mathbf{b}_{m}$ is the bias vector for the $m^{th}$ hidden layer. Finally, the output layer, which is subsequent to the hidden layers, generates the prediction of user $u_i$'s satisfaction with item $v_j$ as:
	\begin{equation}
	{\hat{y}_{i j}=g\left(\mathbf{W}_{o}^{\top} \mathbf{h}_{M}+\mathbf{b}_{o}\right)}
	\end{equation}
\noindent where $\mathbf{W}_{o}$ and $\mathbf{b}_{o}$ are output layer's weight matrix and bias vector. Activation function $g(\cdot)$ varies according to different prediction tasks, such as Sigmoid function for app installation prediction (regression)~\cite{cheng2016wide}, and Softmax for buy-or-not prediction (classification)~\cite{tan2016improved}. By minimizing the loss between predictions and labels, the DLRS updates the parameters of all embeddings as well as neural networks through back-propagation.

\begin{figure*}[t]
	\centering
	\includegraphics[width=156mm]{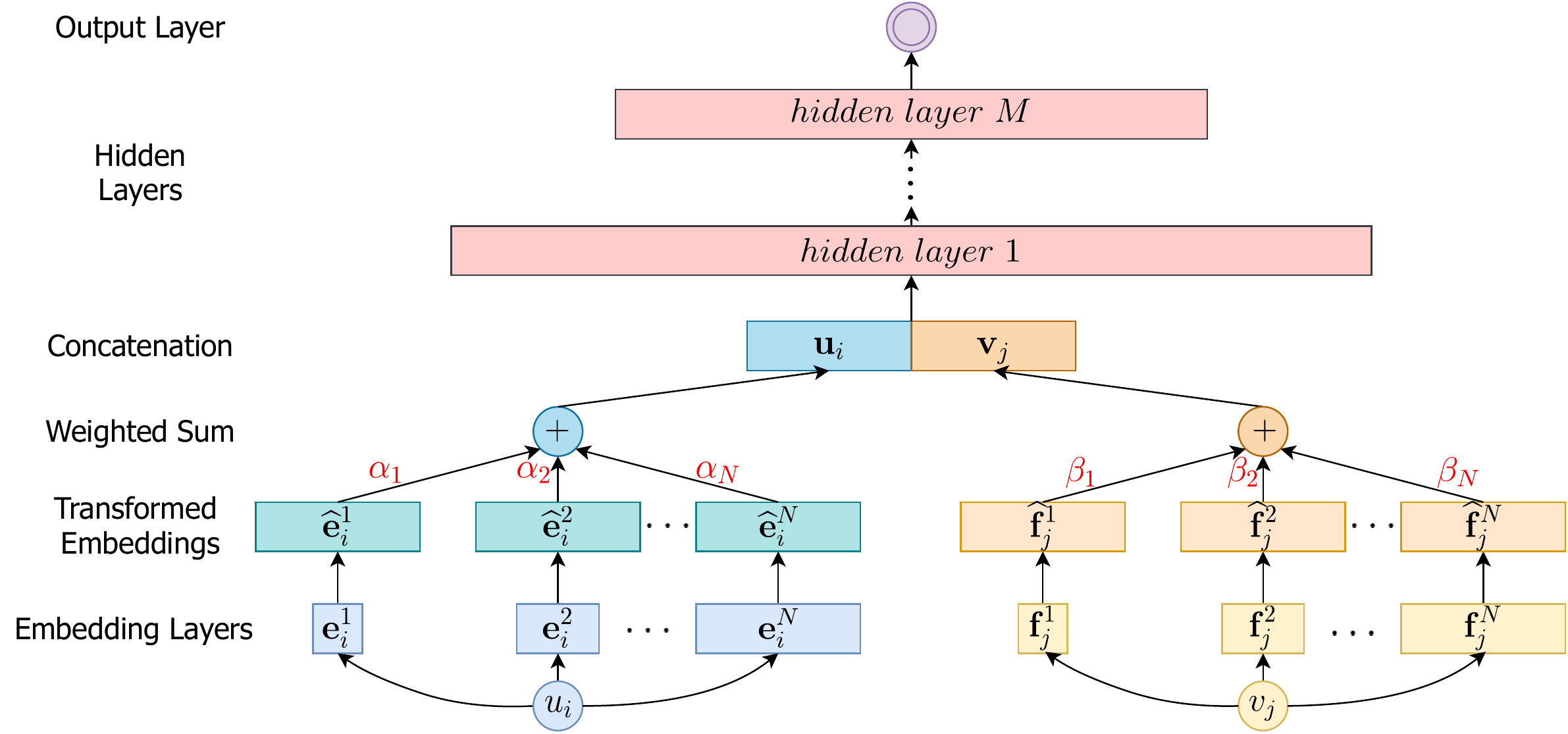}
	\caption{The Enhanced DLRS architecture.}
	\label{fig:Fig5_DLRS1}
\end{figure*}

\subsection{The Controller}
\label{sec:Controller}
Inspired by the observation in Figure~\ref{fig:Fig1_example}, shorter embeddings perform better if popularity is small and longer embeddings perform better when popularity is large. Therefore, it is highly desired to use different embedding sizes for different users and items. Given a large number of users/items and the dynamic nature of their popularity, it is hard, if possible, to determine embedding sizes manually.   



To address this challenge, we propose an AutoML based approach to automatically determine the embedding sizes. To be specific, we design two \textit{controller} networks that decide the embedding sizes for users and items, separately. Figure~\ref{fig:Fig5_controller} illustrates the controller network architecture. For a specific user/item, the controller's input consists of two parts: (i) the \textit{current popularity} of the user/item; and (ii) the \textit{contextual information} such as the previous hyperparameters and the corresponding loss the user/item obtains. This contextual information can be viewed as signals that measure whether hyperparameters assigned to the user/item previously work well. In other words, if they work well, the new hyperparameters generated for this time should be somewhat similar. The controller takes above-mentioned inputs, transforms them via several layers of fully-connected networks, and then generates hierarchical feature representations. The output layer is a Softmax layer with $N$ output units. In this work, we use $\{\alpha_1, \alpha_2, \ldots, \alpha_N \}$ to denote the $N$ output units for the controller of users and utilize $\{\beta_1, \beta_2, \ldots, \beta_N \}$ for items. The $n^{th}$ unit denotes the probability to select the $n^{th}$ embedding space. The embedding space is automatically selected as the one corresponding to the largest probability. It is formally defined as: 
	\begin{equation}
	\label{equ:Hardselection1}
	\arraycolsep=1.2pt\def\arraystretch{1.6}
	\begin{array}{l}
	\mathbf{u}_i = \mathbf{\widehat{e}}_k^{\,*},\;\;\; where \;\; k = \max_{\ell \in \{1,2,\ldots,N\}} \alpha_\ell\\
	\mathbf{v}_j = \mathbf{\widehat{f}}_j^{\,*},\;\;\; where \;\; k = \max_{\ell \in \{1,2,\ldots,N\}} \beta_\ell \\
	\end{array}
	\end{equation}
With the controller, the task of the embedding dimensionality search reduces to optimize the controllers' parameters, so as to automatically generate suitable $\alpha_n$ or $\beta_n$ according to the popularity of a user/item.

\subsection{Soft Selection}
\label{sec:Relaxation}
Eq.~(\ref{equ:Hardselection1}) performs a hard selection on the embedding spaces. In other words, each time, we only select one embedding space with the largest probability from the controller. This hard selection makes the whole framework not end-to-end differentiable. To tackle this challenge, in this work, we choose a soft selection. For $u_i$, its embedding $\mathbf{u}_i$ is a weighted sum of $\{\mathbf{\widehat{e}}_i^1,\cdots,\mathbf{\widehat{e}}_i^N\}$ where the weight of $\mathbf{\widehat{e}}_i^n$ is the corresponding probability $\alpha_n$ from the controller. Therefore, the representations of the user and item can be reformulated as: 
	\begin{equation}
	\begin{aligned}
	\label{equ:WeightedSum}
	\mathbf{u}_i &= \frac{1}{N} \sum_{n=1}^{N} \alpha_n\cdot\mathbf{\widehat{e}}_i^n\\
	\mathbf{v}_j &= \frac{1}{N} \sum_{n=1}^{N} \beta_n\cdot\mathbf{\widehat{f}}_j^n\\
	\end{aligned}
	\end{equation}
With soft selection, the enhanced DLRS is end-to-end differentiable, and we illustrate the whole augmented DLRS architecture in Figure~\ref{fig:Fig5_DLRS1} where we add the transformed embedding layer which performs soft selection of embedding spaces and the selection process is determined by two controllers for users and items, respectively.

\subsection{An Optimization Method}
\label{sec:Optimization}
In this subsection, we investigate the optimization of the proposed framework. With the soft selection, the optimization task is to jointly optimize the parameters of the DLRS, say $\mathbf{W}$, and the parameters of controllers, say $\mathbf{\Theta}$. Since our framework is end-to-end differentiable, inspired by the concept of differentiable architecture search (DARTS) techniques~\cite{liu2018darts}, we adapt a DARTS based optimization for the AutoEmb framework, which updates $\mathbf{W}$ and $\mathbf{\Theta}$ by optimizing the training loss $\mathcal{L}_{train}$ and the validation loss $\mathcal{L}_{val}$ through gradient descent, respectively. Note that both training and validation losses are determined not only by the parameters $\mathbf{W}$ of DLRS, but also the parameters $\mathbf{\Theta}$ of the controller.

The goal for embedding dimensionality search is to find optimal parameters $\mathbf{\Theta}^*$ that minimizes the validation loss $\mathcal{L}_{val} (\mathbf{W}^*,\mathbf{\Theta}^*)$, where the parameters $\mathbf{W}^*$ of DLRS are obtained by minimizing the training loss $\mathbf{W}^* = \arg\min_\mathbf{W} \mathcal{L}_{train} (\mathbf{W}, \mathbf{\Theta}^*)$. This a bilevel optimization problem~\cite{maclaurin2015gradient,pedregosa2016hyperparameter,pham2018efficient}, where $\mathbf{\Theta}$ is the the upper-level variable and $\mathbf{W}$ is the lower-level variable: 
	\begin{equation}
	\begin{aligned}
	\label{equ:bilevel}
	\min_\mathbf{\Theta} \; &\mathcal{L}_{val} \big(\mathbf{W}^*(\mathbf{\Theta}),\mathbf{\Theta}\big)\\
	s.t. \; & \mathbf{W}^*(\mathbf{\Theta}) = \arg\min_\mathbf{W} \mathcal{L}_{train} (\mathbf{W}, \mathbf{\Theta}^*)
	\end{aligned}
	\end{equation}
\noindent Optimizing $\mathbf{\Theta}$ is time-consuming due to the expensive inner optimization of $\mathbf{W}$. Therefore, we leverage the approximation scheme as DARTS: 
	\begin{equation}
	\begin{aligned}
	\label{equ:approximation}
	&\nabla_\mathbf{\Theta} \;\mathcal{L}_{val} \big(\mathbf{W}^*(\mathbf{\Theta}),\mathbf{\Theta}\big)\\
	\approx &\nabla_\mathbf{\Theta} \;\mathcal{L}_{val} \big(\mathbf{W} - \xi \nabla_{\mathbf{W}}\mathcal{L}_{train} (\mathbf{W}, \mathbf{\Theta}),\mathbf{\Theta}\big)
	\end{aligned}
	\end{equation}

\begin{algorithm}[t]
	\caption{\label{alg:DARTS} DARTS based Optimization for AutoEmb.}
	\raggedright
	{\bf Input}: the user-item interactions and the corresponding ground-truth labels\\
	{\bf Output}: well-learned DLRS parameters $\mathbf{W}^*$; well-learned controller parameters $\mathbf{\Theta}^*$\\
	\begin{algorithmic} [1]
		\WHILE{not converged}
		\STATE Sample a mini-batch of validation data from previous user-item interactions
		\STATE Update $\mathbf{\Theta}$ by descending $\nabla_\mathbf{\Theta} \;\mathcal{L}_{val} \big(\mathbf{W} - \xi \nabla_{\mathbf{W}}\mathcal{L}_{train} (\mathbf{W}, \mathbf{\Theta}),\mathbf{\Theta}\big)$ \\($\xi = 0$ for first-order approximation)\\
		\STATE Collect a mini-batch of training data
		\STATE Generate $\alpha_n$, $\beta_n$ via collectors with current parameters $\mathbf{\Theta}$
		\STATE Generate predictions via DLRS with current parameters $\mathbf{W}$ as well as $\alpha_n$ and $\beta_n$
		\STATE Evaluate the predictions and record the performance
		\STATE Update $\mathbf{W}$ by descending $\nabla_{\mathbf{W}}\mathcal{L}_{train} (\mathbf{W}, \mathbf{\Theta})$
		\ENDWHILE
	\end{algorithmic}
\end{algorithm}

\noindent where $\xi$ is the learning rate for updating $\mathbf{W}$. The approximation scheme estimates $\mathbf{W}^*(\mathbf{\Theta})$ by updating $\mathbf{W}$ one training step, which avoids completely optimizing $\mathbf{W}^*(\mathbf{\Theta}) = \arg\min_\mathbf{W}$ $ \mathcal{L}_{train} (\mathbf{W}, \mathbf{\Theta}^*)$ to convergence. The first-order approximation with $\xi = 0$ can even lead to some speed-up, but empirically worse performance. It is worth to note that, different from DARTS on computer version tasks, there is no \textit{deriving discrete architecture} stage, during which DARTS generates a discrete neural network architecture by selecting the most likely operation according to the softmax probabilities. This is because the popularity of users/items is highly dynamic with new user-item interactions occur, which prohibit us from selecting a particular embedding dimension for a user/item.

\begin{small}
	\begin{table}[t]
		\centering
		\caption{Statistics of the datasets.}
		\label{table:statistics}
		\begin{tabular}{c|cc|cc|cc}
			\hline\hline
			Object && Movielens-20m&& Movielens-latest&& Netflix Prize\\\hline\hline
			\# user && 138,493   && 283,228&& 480,189 \\\hline
			\# item && 27,278 && 58,098&& 17,770 \\\hline
			\# interaction && 20,000,263&& 27,753,444&& 100,480,507\\\hline
			\# rating && 1$\sim$5 && 1$\sim$5 && 1$\sim$5 \\\hline
		\end{tabular}
	\end{table}
\end{small}

\begin{table*}[]
	\centering
	\caption{Performance comparison of different embedding selection methods}
	\label{table:result1}
	\begin{tabular}{|c|c|c|c|c|c|c|}
		\hline
		\multirow{2}{*}{Dataset} & \multirow{2}{*}{Task} & \multirow{2}{*}{Metrics} & \multicolumn{4}{c|}{Methods} \\ \cline{4-7} 
		&  &  & FSE & SAM & DARTS & AutoEmb \\ \hline\hline
		\multirow{4}{*}{\begin{tabular}[c]{@{}c@{}}Movielens\\ -20m\end{tabular}} & \multirow{2}{*}{Regression} & MSE Loss  & 0.1840$\pm$0.0003 & 0.1819$\pm$0.0002 & 0.1812$\pm$0.0003 & \textbf{0.1803$\pm$0.0002} \\ \cline{3-7} 
		&  & Accuracy & 0.7211$\pm$0.0003 & 0.7239$\pm$0.0002 & 0.7245$\pm$0.0002 & \textbf{0.7259$\pm$0.0003}  \\ \cline{2-7} 
		& \multirow{2}{*}{Classification} & CE Loss & 1.1464$\pm$0.0006 & 1.1423$\pm$0.0002 & 1.1416$\pm$0.0003 & \textbf{1.1395$\pm$0.0005} \\ \cline{3-7} 
		&  & Accuracy & 0.4943$\pm$0.0003 & 0.4958$\pm$0.0005 & 0.4970 $\pm$0.0002 & \textbf{0.4982$\pm$0.0002} \\ \hline\hline
		\multirow{4}{*}{\begin{tabular}[c]{@{}c@{}}Movielens\\ -latest\end{tabular}} & \multirow{2}{*}{Regression} & MSE Loss & 0.1809$\pm$0.0002 & 0.1803$\pm$0.0003 & 0.1797$\pm$0.0002 & \textbf{0.1790$\pm$0.0001 }\\ \cline{3-7} 
		&  & Accuracy & 0.7260$\pm$0.0003 & 0.7275$\pm$0.0001 & 0.7280$\pm$0.0002 & \textbf{0.7287$\pm$0.0002} \\ \cline{2-7} 
		& \multirow{2}{*}{Classification} & CE Loss & 1.1257$\pm$0.0003 & 1.1249 $\pm$0.0002& 1.1242$\pm$0.0002 & \textbf{1.1233 $\pm$0.0002}\\ \cline{3-7} 
		&  & Accuracy & 0.5049 $\pm$0.0002& 0.5062$\pm$0.0002 & 0.5071$\pm$0.0001 & \textbf{0.5079$\pm$0.0002} \\ \hline\hline
		\multirow{4}{*}{\begin{tabular}[c]{@{}c@{}}Netflix\\ Prize\end{tabular}} & \multirow{2}{*}{Regression} & MSE Loss & 0.1821$\pm$0.0003 & 0.1812$\pm$0.0001 & 0.1807$\pm$0.0001 & \textbf{0.1779$\pm$0.0002} \\ \cline{3-7} 
		&  & Accuracy & 0.7290$\pm$0.0002 & 0.7302$\pm$0.0002 & 0.7309$\pm$0.0003 & \textbf{0.7316$\pm$0.0001} \\ \cline{2-7} 
		& \multirow{2}{*}{Classification} & CE Loss & 1.1102$\pm$0.0003 & 1.1092$\pm$0.0001 & 1.1085$\pm$0.0003 & \textbf{1.1076$\pm$0.0002} \\ \cline{3-7} 
		&  & Accuracy & 0.5096$\pm$0.0003 & 0.5109$\pm$0.0003 & 0.5119$\pm$0.0001 & \textbf{0.5127$\pm$0.0002} \\ \hline
	\end{tabular}
\end{table*}

We present our DARTS based optimization algorithm in Algorithm \ref{alg:DARTS}. In each iteration, we first update controllers' parameters upon validation set collected from previous user-item interactions (line 2-3), then we collect a new mini-batch of user-item interactions as training data (line 4); next we produce hyper-parameters $\alpha_n$ and $\beta_n$ via collectors with its current parameters for the training examples(line 5); then we make predictions via DLRS with its current parameters and the assistance of hyper-parameters (line 6); next we evaluate the prediction performance and record it (line 7); and finally, we update DLRS's parameters.

It is worth to note that, in the batch-based streaming recommendation setting, the optimization process follows an ``evaluate, train, evaluate, train..." fashion~\cite{chang2017streaming}. In other words, we always continuously collect new user-item interaction data; when we have a full mini-batch of examples, we first make predictions based on our AutoEmb framework with its current parameters, and evaluate the performance of the predictions and record it; then we update the parameters of AutoEmb by minimizing the loss between the predictions and ground truth labels; next we collect another mini-batch of user-item interactions and perform the same process. Therefore, there is no pre-split validation set and test set. In other words, \textbf{(i)} to calculate $\mathcal{L}_{val}$, we sample a mini-batch of previous user-item interactions as the validation set; \textbf{(ii)} there is no independent test stage, during which we fix all the parameters and evaluate the proposed framework on examples in the pre-split test set; and (iii) following the streaming recommendation setting in~\cite{chang2017streaming}, we also have \textit{offline parameter estimation} and \textit{online inference} stages, where we use historical user-item interactions to pre-train the AutoEmb's parameters in the offline parameter estimation stage, and then we launch the AutoEmb online and continuously update the AutoEmb parameters in the online inference stage. In other words, AutoEmb's parameters are updated in both stages following the Algorithm~\ref{alg:DARTS}.
	\section{Experiments}
\label{sec:experiments}

In this section, we conduct extensive experiments to evaluate the effectiveness of the proposed AutoEmb framework. We mainly focus on two questions: (i) how the proposed framework performs compared to representative baselines; and (ii) how the controller contributes to the performance. We first introduce experimental settings. Then we seek answers to the above two questions. Finally, we study the impact of important parameters on the performance of the proposed framework.

\subsection{Datasets}
\label{sec:Datasets}
We evaluate our method on widely used dataset: Movielens-20m\footnote{https://grouplens.org/datasets/movielens/20m/}, Movielens-latest\footnote{https://grouplens.org/datasets/movielens/latest/} and Netflix Prize data\footnote{https://www.kaggle.com/netflix-inc/netflix-prize-data}.  Some key statistics of the datasets are shown in Table \ref{table:statistics}. For each dataset, we use 70\% user-item interactions for offline parameter estimation and the other 30\% for online learning. To demonstrate the effectiveness of our framework in the embedding selection task, we eliminate other contextual features, e.g., users' age and items' category, to exclude the influence of other features, but it is straightforward to incorporate them into the framework for better recommendations. 

\subsection{Implement Details}
\label{sec:architecture}
Next we detail the architecture of DLRS and controllers. For DLRS, (i) embedding layer: we select $N=3$ sizes of embedding dimension $[2,16,128]$, thus dimension of transformed embeddings is $128$. We concatenate the three embeddings of each user/item, which significantly improves the embedding lookup speed; (ii) hidden layer: we have two hidden layers with the size $256\times512$ and $512\times512$; (iii) output layer: we do two types of tasks, for rating regression task, the output layer is $512\times1$, and for rating classification task, the output layer is $512\times5$ with Softmax activation, because there are 5 classes of ratings. For controllers, (i) input layer: the input feature size is 38; (ii) hidden layer: we have two hidden layers with the size $38\times512$ and $512\times512$; (iii) output layer: the shape is $512\times3$ with Softmax activation to generate the weights of $N=3$ sizes of embeddings. The batch-size is 500. The learning rate for DLRS and controllers are $0.01$ and $0.001$, respectively. For the parameters of the proposed framework, we select them via cross-validation. Correspondingly, we also do parameter-tuning for baselines for a fair comparison. We will discuss more details about parameter selection for the proposed framework in the following subsections. 

\subsection{Evaluation Metrics}
\label{sec:metrics}
We conduct two types of tasks to demonstrate the effectiveness of the AutoEmb framework. For the regression task, we first binarize ratings to $\{0,1\}$, and then train the framework via minimizing the mean-squared-error (MSE) loss. The performance can be evaluated by MSE Loss and accuracy (we use 0.5 as threshold to assign the labels). For the classification task, the ratings 1$\sim$5 are viewed as 5 classes, and the framework is trained by minimizing the cross-entropy loss (CE Loss). The performance is measured by cross-entropy and accuracy.

\begin{figure*}[t]
	\centering
	\includegraphics[width=176mm]{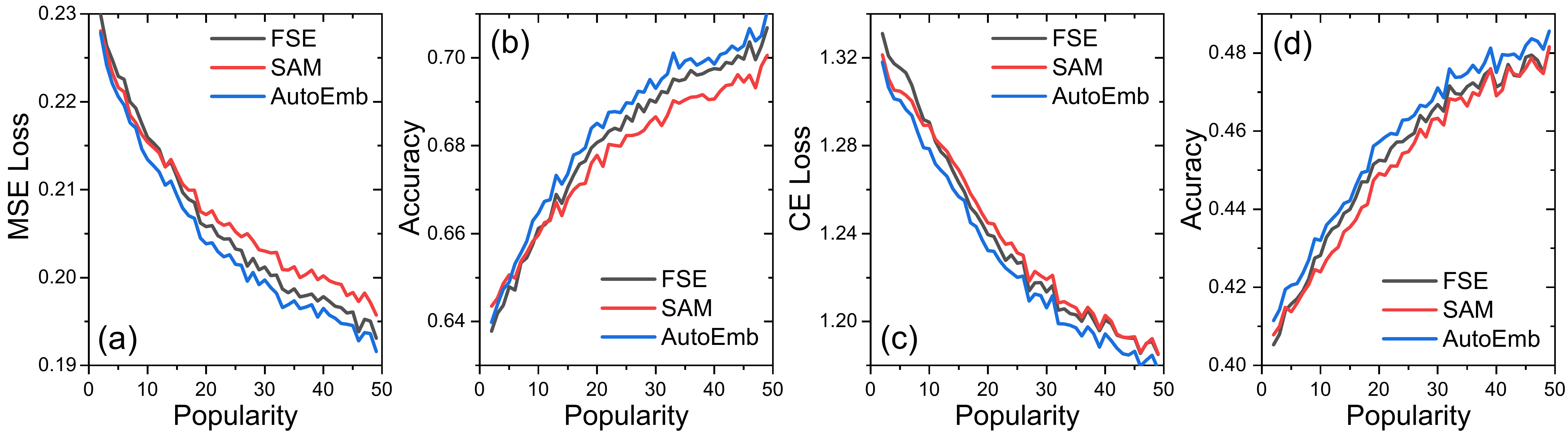}
	\caption{Performance with different popularities.}
	\label{fig:Fig7_popularity}
\end{figure*}

\subsection{Overall Performance in the Online Stage}
\label{sec:Overall}
We compare the proposed framework with the following representative baseline methods:
\begin{itemize}[leftmargin=*]
	\item Fixed-size Embedding (\textbf{FSE}): In this baseline, we assign a fixed embedding size for all the users/items. For a fair comparison, we set the embedding size as $146 = 2+16+128$. In other words, it occupies the same embedding space memory as AutoEmb.
	\item Supervised Attention Model (\textbf{SAM}): This baseline has the exact same architecture with AutoEmb, while we simultaneously update the parameters of DLRS and controllers on the same batch of training data, via an end-to-end supervised learning manner. 
	\item Differentiable architecture search (\textbf{DARTS}): This baseline is a standard DARTS method, which trains $N=3$ real-valued weights for the three types of embedding dimensions.
\end{itemize}

 It is worth to note that, Neural Input Search model~\cite{joglekar2019neural} and Mixed Dimension Embedding model~\cite{ginart2019mixed} cannot be applied in the streaming recommendation setting, because they assume that the popularities of users/items are pre-known and fixed, and then assign highly-popular users/items with large embedding dimensions. However, in real-world streaming recommender systems, the popularities are not pre-known but highly dynamic.

The overall results of the online stage are shown in Table~\ref{table:result1}. We make the following observations:
(i) SAM performs better than FSE, since SAM assigns attention weights on embeddings with different dimensions according to popularity, while FSE has a fixed embedding dimension for all the users/items. These results demonstrate that recommendation quality is indeed related to the popularity of users/items, and introducing different embedding dimensions and adjusting the weights on them according to popularity can boost the recommendation performance.
(ii) DARTS outperforms SAM, because AutoML models like DARTS update controller's parameters on the validation set, which can enhance the generalization, while end-to-end models like SAM  update the parameters of controllers simultaneously with DLRS, on the same batch training data, which may lead to overfitting. These results validate the effectiveness of AutoML techniques over conventional supervised learning in recommendations.
(iii) Our proposed model AutoEmb has better performance than standard DARTS model. DARTS separately trains $N=3$ real-valued weights for each user/item on the three types of embedding dimensions. These weights of a specific user/item may not be well-trained because of the limited interactions of this user/item. The controller of AutoEmb can incorporate huge amounts of user-item interactions and capture the important characteristics from them. Also, the controller has an explicit input of popularity, which may assist the controller to learn the dependency between popularity and embedding dimensions, which DARTS cannot. These results demonstrate the necessity of developing a controller rather than only real-valued weights.
(iv) After the offline parameter estimation stage, most users/items in the online stage have already become very popular. In other words, AutoEmb has stable improvement for popular users/items. AutoEmb has even more significant enhancement in the early training stage, and we will discuss that in the following sections.

To sum up, we can draw an answer to the first question: the proposed framework outperforms representative baselines on different datasets with different metrics. These results prove the effectiveness of the AutoEmb framework.

\subsection{Performance with Popularity}
\label{sec:PerformancePopularities}
Now, we will investigate whether the proposed controller can generate proper weights according to various popularity. Thus, we compare FSE without a controller, SAM with a supervised-attentive controller, and AutoEmb with an AutoML based controller. The results on Movielens-20m dataset are shown in Figure~\ref{fig:Fig7_popularity}, where $x$-axis is popularity and $y$-axis corresponds to performance, we omit the similar results on other datasets due to the limited space. 

\begin{figure}[t]
	\centering
	\includegraphics[width=81mm]{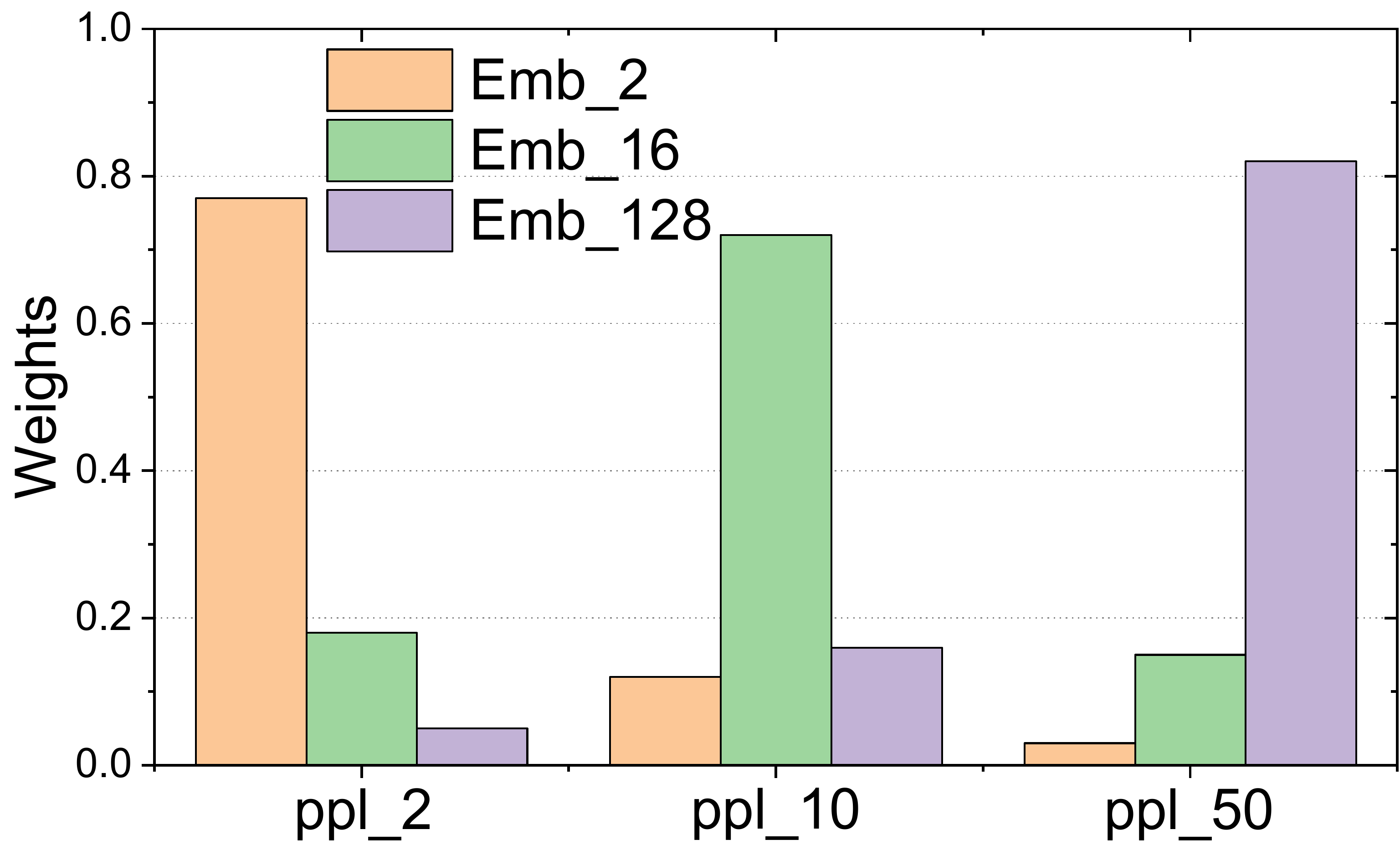}
	\caption{The weights on different embedding dimensions with different popularities (ppl).}
	\label{fig:Fig8_distribution}
	\vspace{-3mm}
\end{figure}

We make following observations: (i) When popularity is small, FSE performs worse than SAM and AutoEmb. This is because larger embeddings need sufficient data to be well learned. Smaller embeddings with fewer parameters can quickly capture some high-level characteristics, which can help the cold-start predictions.  (ii) With the increase of popularity, FSE outperforms SAM. This result is interesting but instructive, and the reason may be that, SAM's controller overfits to a small number of training examples, which leads to suboptimal performance. On the contrary, AutoEmb's controller was trained on validation set, which improves its generalization. This reason is also be validated in the following subsection. (iii) AutoEmb always outperforms FSE and SAM, which means the proposed framework is able to automatically and dynamically adjust the weights on embeddings with different dimensions according to the popularity. (iv) To further probe the weights generated by the AutoEmb's controller according to popularity, we draw the distribution of weights for various popularity in Figure \ref{fig:Fig8_distribution}. We can observe that, the distribution shows on the small embeddings for small popularity, and shows on larger embeddings with the increase of popularity. This observation validates our above analysis. 

In summary, we can answer the second question: the controller of AutoEmb can produce reasonable weights for different popularity via an automated and dynamic manner.

\begin{figure*}[t]
	\centering
	\includegraphics[width=176mm]{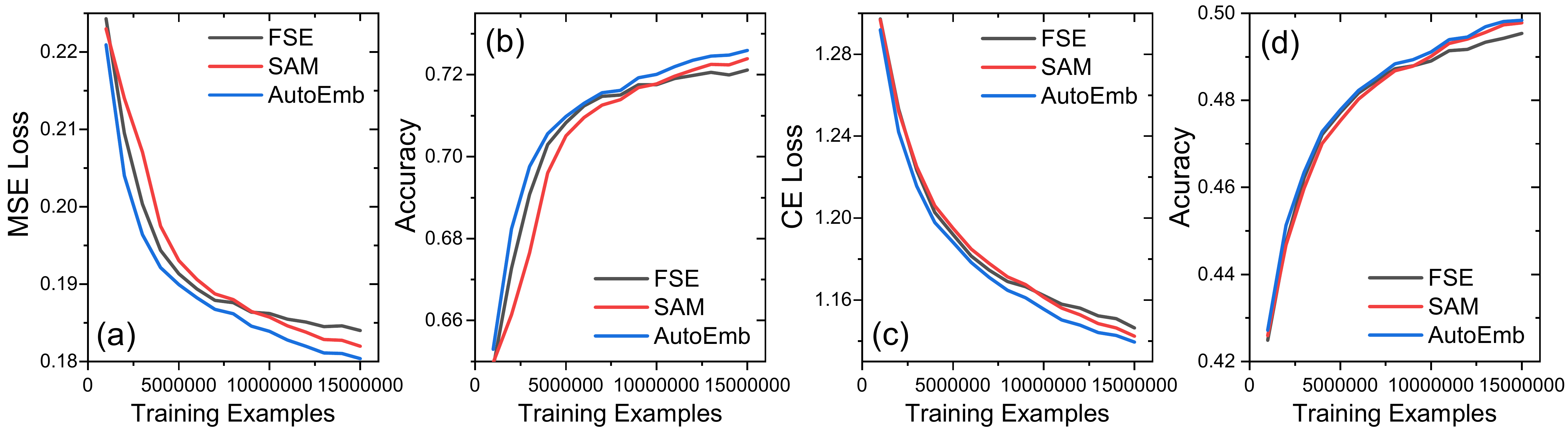}
	\caption{The offline parameter estimation process.}
	\label{fig:Fig9_data}
\end{figure*}

\subsection{Performance with Data}
\label{sec:PerformanceData}
Training deep learning based recommender systems typically requires a large amount of user-item interaction data. Our proposed AutoEmb framework introduces an additional controller network as well as some additional parameters in the DLRS, which may make it hard to be well trained. We show the optimization process in Figure \ref{fig:Fig9_data}, where $x$-axis is the number of training examples, and $y$-axis corresponds to the performance. 

It can be observed: (i) In the early training stage, SAM performs worst since its controller overfits to the insufficient training examples, which is also validated in Section~\ref{sec:PerformancePopularities}. (ii) The overfitting problem of SAM gradually is mitigated with more data coming, and SAM outperforms FSE, which validates the necessity of weights on various embedding dimensions. (iii) AutoML outperforms SAM and FSE in the whole training process. Especially, it can significantly boost the early training stage with insufficient training examples.

	\section{Related Work}
\label{sec:related_work}
In this section, we briefly review works related to our study. In general, the related work can be grouped into the following categories.

The first category related to this paper is deep learning based recommender system, which is able to effectively capture the non-linear and non-trivial user-item relationships, and enables the codification of more complex abstractions as data representations in the higher layers~\cite{zhang2017deep}. 
In recent years, a series of neural recommendation models based on deep learning techniques have been proposed with evident performance lifting. 
He et al.\cite{he2017neural} firstly proposed Neural Collaborative Filtering~(NCF) which utilizes a dual neural network to represent a two-way interaction between user preferences and items features. 
Guo et al.\cite{guo2017deepfm} proposed an end-to-end model named DeepFM to integrate factorization machine and Multilayer Perceptron~(MLP) seamlessly. 
However, some situations would be complex. So the side or extra information could be a booster in improving the performance. 
Xu et al.\cite{xu2016tag}\cite{xu2017tag} fused tag annotations into personalized recommendation and proposed Deep Semantic Similarity based Personalized Recommendation~(DSPR). 
Ali et al.\cite{elkahky2015multi} designed Multi-View Deep Neural Network~(MV-DNN) which could model the interactions among users and items with multiple domains. 
Besides simple expression or transformation of MLP, there are also some other typical models with deep learning methods. 
Suvash et al.\cite{sedhain2015autorec} introduced AutoRec along with item-based and user-based one to learn the lower-dimension feature representations of users and items. 
Lei et al.\cite{lei2016comparative} utilized CNN in image recommendation. Zheng et al.\cite{zheng2017joint} used two parallel CNNs to model user and item features from review texts. 
Hidasi et al.\cite{hidasi2015session} firstly proposed a session-based recommendation model named GRU4Rec to model the sequential influence of items' transition. Lee et al.\cite{lee2016quote} proposed a hybrid model that integrates RNNs with CNNs for quotes recommendation. 
However, most of these works focus on designing sophisticated neural network architectures, while have not paid much attention to the embedding layers.

The second category is about AutoML for Neural Architecture Search~(NAS), which has raised much attention since \cite{zoph2016neural}, which adopts reinforcement learning approach with recurrent neural network~(RNN) to train a large number of candidate models for convergence. Due to the high cost in training, a lot of research focus on proposing novel NAS model with lower hardware resources supporting. One direction is to sample a subset of all model components so that the optimal set could be learned with limited training steps. For instance, ENAS\cite{pham2018efficient} leverages a controller to sample subset of models, and SMASH\cite{brock2017smash} uses a hyper-network to generate weights for sampled networks. DARTS\cite{liu2018darts} and SNAS\cite{xie2018snas} regard the connection as a weight via back-propagation to optimize. Luo et al.\cite{luo2018neural} reflects the neural architectures into an embedding space so that the optimal embedding could be learned and given as feedback to the final architecture. Another direction is to reduce the size of the search space. \cite{real2019regularized, zhong2018practical, liu2018progressive, cai2018path} propose searching convolution cells which could be stacked repeatedly. Zoph et al.\cite{zoph2018learning} developed NASNet architecture which use a transfer learning setting to show smaller datasets could perform better than larger datasets. MNAS\cite{tan2019mnasnet} proposed a hierarchical convolution cell block which could learn different structures. 
NAS has been widely used in different tasks, such as image classification\cite{real2017large}, natural language understanding\cite{chen2020adabert}, etc. Joglekar\cite{joglekar2019neural} firstly utilized NAS into large scale recommendation models and proposed a novel type of embedding named as Multi-size Embedding~(ME). However, it cannot be applied in the streaming recommendation setting, where the popularity is not pre-known but highly dynamic.
	\section{Conclusion}
\label{sec:conclusion}
In this paper, we propose a novel framework AutoEmb, which aims to select different embedding dimensions according to user's/item's popularity in an automated and dynamic manner. In practical streaming recommender systems, due to the huge amounts of users/items and the dynamic nature of their popularity, it is hard, if possible, to manually select different dimensions for different users/items. Therefore, we proposed an AutoML based framework to automatically select from different embedding dimensions. To be specific, we first augment a widely used DLRS architecture to enable it to accept various embedding dimensions, then we propose a controller network, which could automatically select embedding dimensions for a specific user/item according to its current popularity. We evaluate our framework with extensive experiments based on widely used benchmark datasets. The results show that (i) our framework can significantly improve the recommendation performance with highly dynamic popularity; and (ii) the controller trained via an AutoML manner can dramatically enhance the training efficiency, especially when interaction data is insufficient.

There are several interesting research directions. First, in addition to automatically determine the embedding dimensions, we would like to investigate the method to automatically design the whole DLRS architecture. Second, we select the embedding dimensions via a soft manner, which makes the framework end-to-end differentiable, but needs a larger embedding space. In the future, we would like to develop an end-to-end framework with hard selection. Third, the framework is quite general to handle categorical features, thus we would like to investigate more applications of the proposed framework. Finally, we would like to develop a framework that can incorporate more types of features in addition to categorical ones. 
	\bibliographystyle{ACM-Reference-Format}
	\bibliography{9Reference} 
	
\end{document}